# Microservices based Linked Data Quality Model for Buildings Energy Management Services

Muhammad Aslam Jarwar, Sajjad Ali and Ilyoung Chong
Department of Information and Communications Engineering
Hankuk University of Foreign Studies, Yongin-si, Korea
{aslam.jarwar, sajjad, iychong}@hufs.ac.kr

*Abstract*— During the production, distribution, and consumption of energy, a large quantity of data is generated. For efficiently using of energy resources other supplementary data such as building information, weather, and environmental data etc. are also collected and used. All these energy data and relevant data is published as linked data in order to enhance the reusability of data and maximization of energy management services capability. However, the quality of this linked data is questionable because of wear and tears of sensors, unreliable communication channels, and highly diversification of data sources. The provision of high-quality energy management services requires high quality linked data, which reduces billing cost and improve the quality of the living environment. Assessment and improvement methodologies for the quality of data along with linked data needs to process very diverse data from highly diverse data sources. Microservices based data-driven architecture has great significance to processes highly diverse linked data with modularity, scalability, and reliability. This paper proposed microservices based architecture along with domain data and metadata ontologies to enhance and assess energy-related linked data quality.

Keywords— Microservices, linked data quality, data quality, energy management services;

## I. Introduction

The energy management applications are using information and communication technologies for dealing with the rapid urbanization and the provision of improved energy management services in order to increase the sustainability of resources, and maximizing the comfortable livings [1]. To use energy resources efficiently and effectively with the consideration of lowering the greenhouse gases, the collected energy data needs to be with sufficient quality for better decision making. However, the collected energy-related data quality is questionable because it has missing quality factors which should need attention in order to improve the services quality. Sometimes the collected energy data are wrongly annotated and mislabeled and also contains gaps and errors. which can mislead in the forecasting the demand and supply of energy, inaccurate billing, missed energy saving or retrofit opportunities.

Nowadays, mostly energy data is published as linked data in order to make it widely accessible and reusable across organizational boundaries by many applications (i.e. energy generation companies, energy distribution companies, appliances production companies, and typically buildings applications HVAC, lighting devices e.t.c). The linked data in generally and specifically in energy-related data needs to be modeled, published, discovered, and integrated with respect to all stakeholders in order to increase data accessibility and usability. The concept of linked data promote the publication of data using web standards, encourage the reusability of data, reduces the redundancy of data and redundant data management efforts, and foster others to add value to the data [2]. The publication of energy data as linked data make it machine-readable, reduces energy data isolation, and accessible with existing web-standard [3].

As there are many stakeholders and variety of data sources sources in energy-eco-cycle, therefore, energy linked data grows very rapidly. Therefore, to increase the energy linked data quality it needs to preprocess and analyzed its quality in many dimensions with modular data preprocessing capabilities. Microservices foster the pre-processing and analysis of data with high accuracy, currency, and modularity [4], [5]. It supports the containerization of data quality dimension measuring procedures with rapid processing, development, and deployment. Microservices are developed around the business and operational capabilities and can be deployed and scaled automatically according to the service requirement of the application. Because the linked data quality dimensions are interrelated, on one hand, to improve some DQ dimensions negatively affect the other dimensions. For example to pre-process the data to increase the semantic accuracy of the data negatively affect the timeliness of data [6].

In this article, we discuss linked data quality dimensions and metrics in generally and particularly with respect to energy linked data. We present the microservices based architecture to process energy linked data in order to check its data quality dimensions and resolve the gaps and errors in the data.

## II. Related work

Microservices pattern is the extension of service-oriented architecture (SOA) [7]. This pattern supports the creation of service from the suite of small services [5], [8], [9]. In the microservices style architecture, the data among the small services are passed through message queue in order to reduce latency and increase the performance. Microservices foster the data processing in a dynamic environment because of its three-dimensional scalability concept [4][10]. To expose the services capability with data quality, microservices style architecture has significance support. In such a dynamic IoT environment and energy management environment, where the applications deal with a variety of data sources from heterogeneous objects and to enhance data quality for the quality of services, microservices based design and implementation architecture is a suitable way [11]–[13].

## III. Linked data quality dimensions and metrics

The data quality is a multi-dimensional concept and it refers that how the data is fit for use in a particular context and services for its consumers [14], [15]. In the literature data quality assessment is categorized into subjective data quality assessment and objective data quality assessment [16], [17]. The subjective data quality assessment influenced by personal experience and behavior. This type of assessment usually conducted by the data producer or data stakeholders. Whereas the objective data quality assessment can be task independent and task dependent. The task dependent data quality assessment procedure focused on the application domain and context on the other hand task independent assessment usually conducted without concentrating on the context of services and application domain as shown in Figure 1. The energy linked data quality assessment falls in the category of objective assessment with respect to the building energy management services. Because here we are considering specific application domain and services such as recommending services [18].

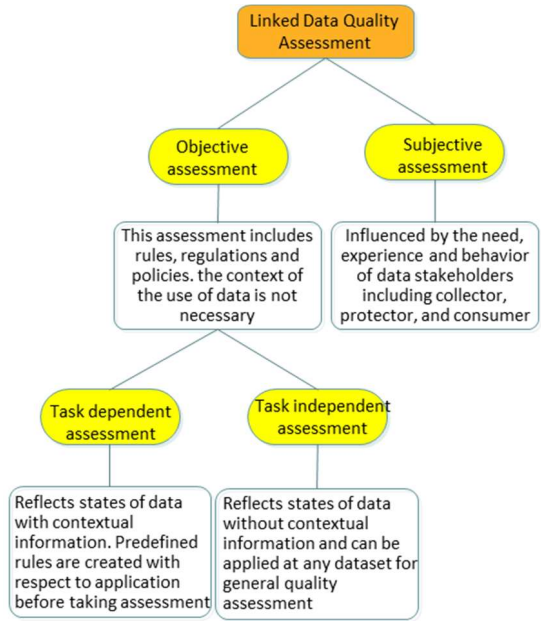

Figure 1. Linked data quality assessment categorization and description

## IV. LINKED DATA QUALITY DIMENSIONS WITH RESPECT TO ENERGY DATA

For the provision of effective and efficient energy management services, we need many types of integrated data with sufficient data quality. The visual excerpt of energy-related data types and their description with attributes is shown in figure 2.

We select and extends the linked data quality dimensions with respect to energy-related linked data datasets. The each linked data quality dimensions set will be discussed with energy data examples in the following sub-sections.

### A. Accessibility data quality dimensions set:

The accessibility dimensions are more related to the linked data infrastructure [19], access rights level such as up to how much interlinks are accessible, and time duration in which data will be accessible. In this accessibility linked data quality dimensions set, the data availability, data interlinking, and the performance of the endpoint where the data is deployed are important to energy-related linked data quality

### B. Intrinsic data quality dimensions set:

The data quality dimensions in this category focused on the data itself rather than the data metadata. Data quality dimensions in this group can be measured by checking the data gaps or missing data, data outliers, and inconsistency in the data. During data processing for improving the results of intrinsic data quality dimensions, the usage of the data at the end applications is not considered. The intrinsic data quality dimensions in the energy-related linked data domain can be measured with data consistency, data semantic accuracy, and data completeness.

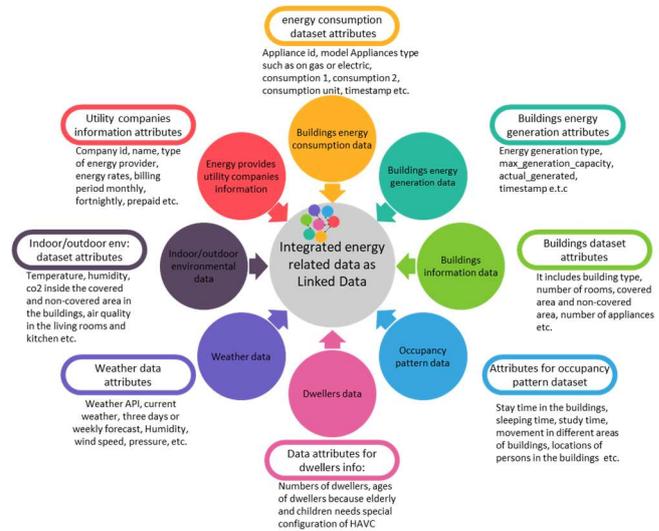

Figure 2. Excerpt view of integrated energy data types and attributes as linked data

### C. RDF level data quality dimensions set:

The data quality dimensions in this category are related to the data modeling and data formatting aspects such as the RDF data model. That how the data and metadata is well legitimated and orchestrated in a compact way, in the document or in the database. So that it can be easy to interpret and understandable across the related application domains and services when it retrieved and used. For example buildings energy management services used many types of energy-related data for the provision of better quality services, for different types of users and stakeholders. Therefore, the data should be understandable and interpretable by all the users and stakeholders. Data quality aspects in this category can be identified by measuring the data interpretability, data interoperability [20], and data compactness. The data quality characteristics of this category dimensions can be improved with the utilization of domain data and metadata ontology [19].

### D. Task dependent data quality dimensions set:

The task dependent linked data quality dimensions measured level remain dynamic and can be changed based on the service or task for which the data quality is measured. For the similar dataset, the data quality can be very good in one service and very bad in the other service. For example, the home appliances usage history dataset usability and reliability can be good to predict energy requirement for the next quarter in a home environment. But not good in order to predict in the commercial environment. The data source provenance, data freshness, and how much the data is usable for the specific service are the more important aspects of the task dependent data quality dimensions set.

## V. MICROSERVICES BASED ARCHITECTURE TO PROCESS ENERGY LINKED DATA QUALITY FOR SERVICES

In order to improve energy-related linked data quality for efficient and effective energy management services, we designed a microservices based modular and layered architecture as shown in figure 3.

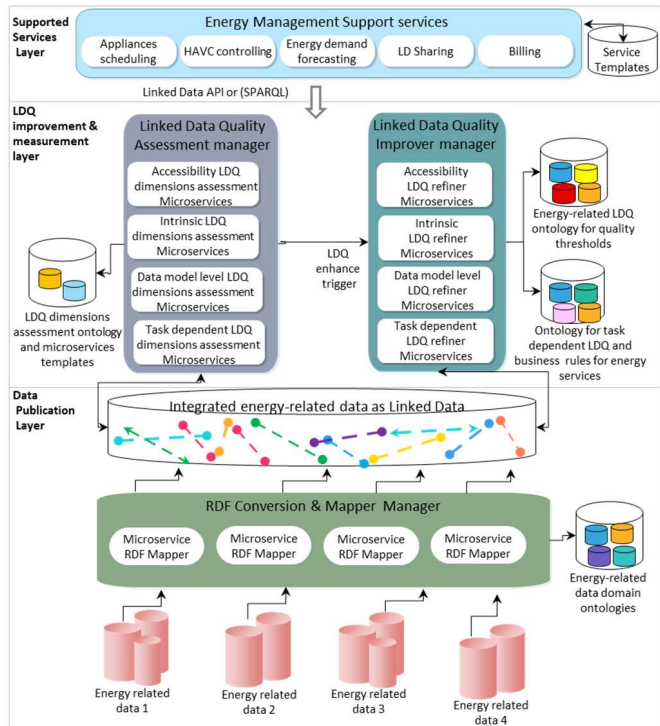

Figure 3. Microservices based linked data quality processing for energy management services

The architecture follows the bottom-up pattern. In the bottom, there are various types of energy-related data sources such as buildings energy consumption data, buildings energy generation data, buildings information data, occupancy pattern data, dwellers information, weather data, indoor/outdoor environmental data and etc. The data sources can be from sensors, actuators, smart meter or 3rd party services APIs. Assume that these energy-related data is not RDF based data, therefore, when the instances of data received. It can be converted and mapped to RDF format, i.e. subject, object, and predicate by using the energy-related data domain ontologies. The conversion, mapping, and annotation to RDF data model will be performed by microservice RDF mapper for each data types. Energy-related data domain ontologies contain the class axioms, instances, object properties and data properties for each type of energy-related data and are designed by the domain expert. The RDF Conversion and Mapper Manager will expose this converted energy-related data as published linked data.

The published energy-related linked data quality will be checked and assessed by the Linked Data Quality Assessment Manager (LDQAM) with respect to four types of data quality dimensions as discussed in the previous sections. The LDQAM is the suite of four types of microservices with respect to each linked data quality dimensions and each type of microservices also contains nanoservices for measuring each dimension in the category. The microservices based LDQAM can easily be extended when required with more data quality dimension types and assessment metrics. In the LDQAM microservices, the accessibility dimensions can be measured with average IRI dereferenceability, average subject dereferenceability by using HTTP status codes (i.e. 404 not found). The intrinsic type dimensions can be measured by the detection of missing or misuse of data property and object properties, incompatible energy measurement units, and environment measurement units. The microservices in the LDQAM are created and configured with microservices templates and assess the quality of linked data by using the assessment ontology. The assessment ontology contains the facts and procedures to measure the linked data quality with defined dimensions against each energy-related data types. When the LDQAM completes the data quality assessment, then it checks whether the quality is sufficient for business objectives such as energy management services as defined in the assessment ontology. If the assessed quality is below the threshold then it triggers the message to the Linked Data Quality Improver Manager (LDQIM).

The linked data quality improver manager is using four types of microservices to refine the linked data quality in four different aspects such as accessibility aspects, intrinsic aspects, data modeling level aspects, and task dependent related aspects. For refining the data quality, microservices are using ETL (Extract, Transform and Load) process along with energy-related data quality thresholds values, and business rules and policies from the two different ontologies as indicated in figure 3. The examples to improve energy-related linked data quality are: in energy-related data some values can be in out-of-range or missing and also contains invalid data properties, invalid dereferenceable links, data duplication, different names for same appliance, different data freshness time alignment, data model conventions, 3rd party data provenance and etc.

## VI. ONTOLOGY FOR ENERGY LINKED DATA QUALITY MODEL

To improve and measure the integrated energy linked data quality, we extended the W3C Linked Data Quality Vocabulary[1] [21]. The excerpt view of the extended ontology is shown in figure 4. In the ontology, the grey circle shows the classes of W3C data quality vocabulary and prefixed with dqv. i.e. *dqv:dimension*, and dqv:*precision*. The extended classes are prefixed with eldv. i.e. *eldv:accessibility*, *eldv: consistency*, *eldv:RDF level or data model level*, and *ldv: tasks dependent*. This ontology also contains concepts for improvement of energy linked data quality with respect to dimensions and improved precision. The data quality precision information is the degree at which the data quality is improved with the improvement techniques and methods.

The linked data quality improvement methods include clustering data interlinking (*eldv:clustering data interlinking*), support vector regression (*eldv:support vector regression*), and association rule mining (*eldv:association rule mining*).

By using the clustering data interlinking method, we can connect the related linked data concepts. The association rule mining method such as a priori can be used to find out the missing object in the triple based on the subject and predicate, and missing subject based on the object and predicate. All the data and logical modeling for the data quality methods are based on the microservices. The linked data quality improvement methods can be developed with microservices modeling independently and deployed in containers for the good performance in order to support and minimize the inversely effect of correlated data quality dimensions, i.e. data freshness vs data accuracy and data usability.

---

[1] https://www.w3.org/TR/vocab-dqv/

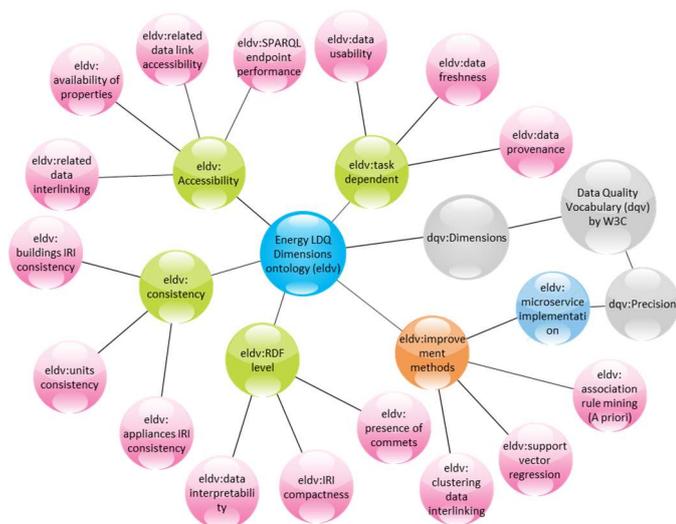

Figure 4. Excerpt view of ontology for energy linked data quality model (grey circle shows the W3C Data Quality Vocabulary and shown with dqv prefix and the extension is prefixed with eldv)

## VII. CONCLUSION

In the production, distribution, and consumption of energy tremendous amount of energy-related data is generated. This generated energy-related data is collected, discovered, and integrated as linked data for the provision of energy management services. Due to wear and tear in sensors and communication links and ambiguity in data schema, the quality of that data is questionable. We need to process that data, perform the data quality assessment and then design the data quality improvement plan and then monitor the data quality in order to provide high-quality energy efficiency management services. This article proposed microservices based linked data quality model to assess and improve the energy-related data with defined data quality dimensions and metrics. In the article, we also defined the energy-related data domain ontology and task dependent linked data quality assessment business rule ontology.

### ACKNOWLEDGMENT

This research was supported by the MSIT (Ministry of Science and ICT), Korea, under the ITRC (Information Technology Research Center) support program (IITP-2018-0-01396) supervised by the IITP (Institute for Information & communications Technology Promotion)